\documentclass[a4paper,11pt]{article}
\usepackage{aaskaiid}
\usepackage{orcidlink}
\usepackage{color}
\title{Probing the Parsec-Scale Dynamical Structure of Ionized Gas in Radio-Quiet AGN with SKA}
\ShortTitle{Parsec-Scale Ionized Gas in RQAGN}

\author[1,2]{Yuki Kudoh \orcidlink{0000-0003-0548-1766}}
\author[3,4]{Satoko Sawada-Satoh \orcidlink{0000-0001-7719-274X}}

\affiliation[1]{Information Technology Center, The University of Tokyo, Kashiwa, Chiba, Japan}
\emailAdd{yuki.kudoh@cc.u-tokyo.ac.jp}
\affiliation[2]{Astronomical Institute, Tohoku University, Sendai, Miyagi, Japan}
\affiliation[3]{Department of Electrical, Electronic and Computer Engineering, Faculty of Engineering \\
 Fukui University of Technology, Gakuen, Fukui-City, Fukui, Japan}
\affiliation[4]{Graduate School of Science, Osaka Metropolitan University, Sugimoto, Sumiyoshi-ku, Osaka, Japan}

\abstract{
We systematically organize the radio-emitting components in radio-quiet active galactic nuclei (RQ AGN), including jets, accretion disk coronae, dust, ionized gas outflows, and circumnuclear star formation.
We present a diagnostic framework for distinguishing these components using spectral turnovers and spectral indices produced by synchrotron self-absorption (SSA) and free-free absorption (FFA), together with brightness temperature and peak frequency.
The central premise is that the observed spectral index and its spatial distribution are not unique source properties unless the observing beam is specified: changing the angular resolution changes the physical scale being sampled and therefore changes the mixture of radio-emitting components.
By exploiting this scale dependence with SKA1-MID and SKA-VLBI, spatially resolved spectral-index mapping will reveal which physical processes dominate from circumnuclear star formation on $\sim$100~pc scales to jets, coronal emission, and compact ionized gas on parsec and sub-parsec scales.
Through multi-frequency continuum imaging and spectral-index mapping, SKA observations will provide a multi-scale physical view of radio-quiet AGN that links radio emission mechanisms to accretion, obscuration, and feedback.

}

\begin{document}
\maketitle

\section{Introduction}
Active Galactic Nuclei (AGN) are celestial objects that host supermassive black holes (SMBHs) at their centers and exhibit diverse radiation processes. The spectral energy distribution from infrared (IR) to X-ray observations has established that their emission arises from processes directly or indirectly related to matter accreting onto the SMBH. Radio classification is quantitatively defined using the flux density ratio R between the radio flux $F_{\rm 5GHz}$ at 5 GHz and the optical flux density in the B-band (effective frequency $\sim 6.8\times 10^{5}$ GHz), as $R = {F_{\rm 5GHz}}/{F_{\rm B}}$.
The standard definition classifies objects with $R > 10$ as radio-loud (RL) and those with $R < 10$ as radio-quiet (RQ) \citep{1989AJ.....98.1195K}. While RL AGN are dominated by synchrotron radiation from powerful relativistic jets, the radio emission mechanism in RQ AGN remains poorly understood.

Understanding the relationship between gas structures and radio emission in AGN requires consideration of multiple structural components including small-scale jets around the accretion disk, disk coronae, ionized gas outflows, and star-forming regions \citep[reviewed by][]{2019NatAs...3..387P}. In RQ AGN particularly, contributions from diverse emission sources are intermixed. Nevertheless, radio continuum emission from each component provides important clues for understanding the spatial distribution of ionized gas structures in AGN, offering significant potential for synergistic interpretation with multi-frequency data.

The Square Kilometre Array (SKA) can provide sensitivity and spatial resolution far exceeding those of conventional facilities, enabling detailed exploration of faint radio sources in RQ AGN. If high spatial resolution multi-band observations can spatially separate individual structures, this will provide a crucial opportunity to understand in detail the physical properties of each emission source. 
In this report, Section 2 presents candidate radio emission sources in the RQ AGN core region and their emission processes, as well as the current understanding of radio emission in radio-quiet AGN. In Section 3, SKA will turn beam-dependent spectral-index changes from a source of ambiguity into a physical diagnostic of scale-dependent emission processes. We discuss the synergies of multi-frequency observations during the SKA era in Section 4.

\section{Radio Continuum Emission in AGN: Structure and Spectral Energy Distribution}

In active galactic nuclei (AGN), multiple structural components—jets, disk coronae, dusty torus or circumnuclear disk, embedded star-forming regions, and ionized gas outflows—can contribute to the observed radio-to-submillimeter emission. Each component has characteristic spatial scales and emission mechanisms, imprinting a distinctive spectral energy distribution (SED). The observed flux density can be expressed as a power law, 
\begin{equation}
F_\nu \propto \nu^{\alpha}, \label{eq:1}
\end{equation}
as a function of frequency $\nu$, where the spectral index $\alpha$ is determined by the underlying radiation process. Here we adopt the convention in which $\alpha$ is negative for optically thin synchrotron emission and positive for optically thick (inverted) spectra. These power-law spectra are modulated by absorption processes at low frequencies: as the optical depth increases below a characteristic frequency, a spectral turnover occurs. Consequently, flat or inverted spectra ($\alpha \geq 0$) are often observed on the low-frequency side. In what follows, we describe the spatial structures and power-law indices associated with each component.

A key point for SKA observations is that a measured spectral index is always beam-dependent. If the beam includes both a steep-spectrum star-forming ring and a flat or inverted compact AGN core, the measured $\alpha$ is a luminosity-weighted mixture; if the same source is observed with a smaller beam, the extended component is resolved out and the inferred $\alpha$ can change even without intrinsic variability. This behavior is not merely an observational nuisance. Multi-resolution $\alpha$-mapping provides a way to assign radio-emission mechanisms to physical scales: $\sim$100~pc beams emphasize circumnuclear star formation and extended outflows, $\sim$10~pc beams isolate the torus/NLR interface and compact jet--ISM interactions, and mas-scale beams test sub-pc SSA/FFA components in the jet base, corona, and RPC ionized skin. Similar scale-dependent interpretations underpin classical radio diagnostics of star formation and compact AGN cores \citep{1992ARA&A..30..575C,2001ApJ...549L.167Y,2003ARA&A..41..117B,2019NatAs...3..387P}.

\subsection{Relativistic Jets and Compact Cores}
AGN jets emit synchrotron radiation, producing a power-law spectrum Eq.\ref{eq:1} with $\alpha=(1-p)/2$ for an electron energy distribution $N(E)\propto E^{-p}$. 
Optically thin jets typically exhibit $\alpha\sim-0.7$ to $-1.0$, whereas compact, optically thick radio cores display flat or inverted spectra ($\alpha\approx 0$) \citep{2001PASJ...53..169K}.

In the case of synchrotron self-absorption (SSA), a homogeneous synchrotron source exhibits a low-frequency spectrum of $S_\nu \propto \nu^{5/2}$. 
VLBI observations of M87 at 22 and 129 GHz reveal a flat core spectrum ($\alpha \gtrsim -0.4$), inconsistent with a single homogeneous SSA component, implying multiple synchrotron components with turnover frequencies straddling $\sim 100\,{\rm GHz}$ \citep{2018A&A...610L...5K}.
Spectral-index maps show a transition from an inverted/flat spectrum ($\alpha \approx +0.2$ to $0$) at the core, to moderately steep values ($\alpha \approx -0.5$ to $-1$) downstream, and ultimately to $\alpha \approx -1.5$, consistent with the Blandford-K\"{o}nigl jet model \citep{2023MNRAS.526.5949N}. 
Detection of a frequency-dependent core shift provides a strong diagnostic of SSA. 
However, whether the SSA framework established for RL AGN can be directly applied to RQ AGN remains uncertain.

Another cause of low-frequency cutoff is free-free absorption (FFA) by foreground thermal plasma. NGC 1052 exemplifies the combination of SSA and FFA: VLBI observations at 5--43 GHz reveal a symmetric two-sided jet, yet only the approaching jet is detected at low frequencies ($\lesssim$8-10 GHz). 
\cite{2001PASJ...53..169K} found an unrealistically inverted spectral index at the core ($\alpha_{8.4-15.4}\approx+3.1$).
\cite{2009AN....330..249S} also reported strong inversion at 15-43 GHz. 
 This cannot be explained by SSA alone and indicates the presence of a dense free-free absorber such as a circumnuclear torus. In a torus geometry, the receding jet traverses a longer path than the approaching jet, resulting in a higher $\nu_{\rm peak}$. 
 Physical conditions of the absorber ($n_e\sim10^5$ cm$^{-3}$, $T_e\sim10^4$ K) have been derived from VLBI spectral maps, confirming that free-free opacity dominates within the innermost parsec scale.

Observationally, distinguishing between the two absorption processes, SSA and FFA, is challenging. In many radio cores, the spectrum rises steeply below a turnover frequency $\nu_{\rm t}$ ($\sim$0.1-10 GHz). A slope $\alpha \lesssim 5/2$, the maximum value attainable for a homogeneous SSA source, is consistent with SSA, whereas a significantly steeper cutoff ($\alpha \gg 5/2$) indicates FFA. Physical parameters can be derived from measurements of the turnover. By measuring the peak flux density $S_m$, peak frequency $\nu_m$, and angular diameter $\theta$ from VLBI spectra, the magnetic field $B$ and brightness temperature $T_b$ can be obtained. Typical cores exhibit $T_b\sim10^9$-$10^{11}$ K and $B\sim10^{-3}$-$1$ G. When FFA dominates, the condition $\tau_{\rm ff}\sim1$ yields the electron column $n_e^2L\propto\tau_{\rm ff} \nu^{2} T_e^{3/2}$.

\subsection{Accretion Disk Corona}

In geometrically thin, standard accretion disks \citep{1973A&A....24..337S}, a hot plasma known as the disk corona forms above the disk and plays a crucial role as a compact source of SSA. The synchrotron critical frequency \citep{2018ApJ...869..114I} is given by
\begin{equation}
\nu_{\rm c} \simeq \frac{3}{2}\gamma^2 \frac{eB}{2\pi m_{\rm e} c}\sin\theta \simeq 4.2\times10^{6}\,B({\rm G})\,\gamma^2 \sin\theta\ {\rm Hz},
\end{equation}
where $\theta$ is the pitch angle between the electron velocity and the magnetic field, $\gamma$ is the electron Lorentz factor and $B$ is the magnetic field strength.
Although low-radiative-efficiency accretion flows (ADAF/RIAF) also exhibit a rising millimeter bump produced by synchrotron self-absorption of thermal electrons, the physical origin differs: in ADAFs, the innermost accretion flow itself becomes optically thick to SSA, whereas in standard disks, the emission originates predominantly from the overlying corona.

Recent millimeter and X-ray observations have revealed a positive correlation between the millimeter luminosity ($L_{\rm mm}$) and the X-ray luminosity ($L_X$) \citep{2018MNRAS.478..399B,2022ApJ...938...87K}. 
This correlation suggests that both components originate from the same coronal plasma.
Millimeter data alone constrain the magnetic field strength ($B$) and source size ($R$) through the synchrotron self-absorption (SSA) turnover, but they leave a degeneracy with respect to the electron energy distribution.
In contrast, reproducing the observed X-ray luminosity assumed to be dominated by inverse Compton scattering requires consistency in the electron number ($N_e$), energy distribution, and seed-photon energy density.

The millimeter luminosity can be expressed as $L_{\rm mm} \sim \eta_{\rm syn}\,U_B\,V$, with $U_B = B^2/8\pi$ and $V \sim 4\pi R^3/3$, where $\eta_{\rm syn}$ is the synchrotron efficiency, $U_B$ the magnetic energy density, and $V$ the emitting volume \citep{2025A&A...701A..41D}.
If the hard X-ray component arises from inverse Compton scattering, its luminosity is approximately
$L_X \sim \eta_{\rm IC}\,U_{\rm ph}\,N_e\,c\,\sigma_T$,
where $U_{\rm ph}$ denotes the seed photon energy density and $N_e$ the electron number.
Requiring a single electron population to simultaneously explain both $L_{\rm mm}$ and $L_X$ breaks the degeneracies present when using either band alone, yielding quantitative and self-consistent constraints on key coronal parameters such as $B$, $R$, and the electron energy distribution.

\subsection{Dusty Ionized Gas Outflows}

Within a few parsecs, the dusty and ionized gas exhibits a multi-phase structure comprising the obscuring torus, the narrow-line region (NLR), and the broad-line region (BLR). 
The dusty material surrounding the nucleus shows anisotropic infrared emission that depends on the line-of-sight column density, explaining the degree of obscuration of the active galactic nucleus as Type 1 and Type 2.
The NLR often contains ionized gas outflows, which appear as blueshifted asymmetric components in the [O\,{\sc iii}]\,$\lambda$5007 emission line.
Spectral decomposition enables the identification of outflow components within the NLR \citep[e.g.][]{2014ApJ...795...30B,2017A&A...601A.143F}.
Furthermore, the ionized gas outflows are spatially associated with polar dust, a radiation-driven mixture of gas and dust lifted from the disk plane \citep{2017ApJ...838L..20H,2019MNRAS.484.3334S,2022ApJ...940...28K}.
This polar dust structure has been directly detected through infrared interferometry, revealing ionized gas outflows located above the BLR and the inner torus \citep{2024A&A...690A..76G}.
Nevertheless, the detailed spatial and dynamical structures of the ionized gas remain uncertain.

The ionized gas can produce either thermal free--free emission, characterized by a nearly flat spectral index ($\alpha \approx -0.1$), or non-thermal synchrotron emission from shock-accelerated electrons ($\alpha \lesssim -0.7$). For example, in NGC~1068, wide-angle ionized outflows traced by [O\,{\sc iii}] emission have faint radio counterparts, suggesting radio emission from shock-driven jet/outflow--ISM interactions \citep[e.g.][]{2004ApJ...613..794G, 2020ApJ...893...33L}.

On sub-parsec scales, the radiation pressure confinement (RPC) framework has been proposed to explain the ionized gas structures in the BLR and NLR \citep{2018MNRAS.474.1970B,2021MNRAS.508..680B}. 
In this model, the balance between AGN radiation pressure and gas pressure naturally leads to a radial gas density profile expressed as
\begin{equation}
    n_e \sim \frac{L_{\mathrm{ion}}}{4\pi r^2 c k_B T},
\end{equation}
where $c$ is the speed of light, $k_B$ is the Boltzmann constant, and $T$ is the ionized gas temperature.
Since this relation gives $n_e \propto r^{-2}$ at fixed $T$, the ionization parameter defined in X-ray studies as
\begin{equation}
\xi = \frac{L_{\mathrm{ion}}}{n_e r^2},
\end{equation}
remains approximately constant with radius (specifically, substituting the above expression yields $\xi \simeq 4\pi c k_B T$, which depends only on the gas temperature).
This self-regulated condition allows for a continuous ionization structure extending from the BLR to the NLR.
In the outer layers of this structure, as the photon density decreases, a high-density ionized skin forms with a typical temperature of $T_e \sim 10^4$~K.
This layer is proposed to be the source of compact, sub-parsec-scale free--free emission \citep{2021MNRAS.508..680B}.

Recently, radiation--hydrodynamic simulations \citep{2023ApJ...950...72K,2024ApJ...977...48K} have demonstrated that radiation-driven ionized winds can naturally develop high-density layers.
Photoionization heating by UV and X-ray photons establishes gas pressure within the outflow, and in regions where radiation pressure and gravity achieve near balance, the flow forms quasi-static layers with number densities of $10^6$~cm$^{-3}$.
These high-density ionized layers, similar to those predicted under the RPC condition, can be dynamically realized in radiation-driven outflows.

A related study by \citet{2026ApJ...998...60W} further demonstrates that ionized gas in the BLR-scale environment can affect observed recombination-line profiles.
By coupling three-dimensional radiation--hydrodynamic simulations with CLOUDY radiative-transfer calculations for NGC~3783, they showed that Br$\gamma$ emission arises from ionized gas near the disk surface, while electron scattering in surrounding diffuse ionized gas can broaden and smooth the intrinsic line profile.
This result indicates that ionized gas produced in radiation-driven AGN environments can influence not only radio free--free opacity and compact continuum emission, but also near-infrared line widths and line shapes.

\subsection{Cold Dust Component in the Circumnuclear Disk}
In the infrared to submillimeter regime ($\sim10^{11}-10^{12}$ Hz), warm dust within the torus (T $\sim 100-300$ K) behaves as a thermal emitter whose spectrum can be approximated by $F_\nu \propto \nu^{2+\beta}$, where $\beta$ is the dust opacity index, typically ranging between 1 and 2.
ALMA observations of NGC 1068 have spatially resolved the dust torus on scales of a few parsecs, revealing that the $100-500$ GHz continuum contains contributions from both thermal dust emission and free-free emission \citep{2016ApJ...823L..12G,2019A&A...632A..61G,2018ApJ...853L..25I}.

The far-infrared to millimeter spectral energy distribution (SED) of galaxies is commonly characterized by a modified blackbody (MBB) model with typical dust temperature of $T_d \sim 20-30$ K and an emissivity index of $\beta \sim 1.5-2$ \citep{2007ApJ...657..810D}.
However, at frequencies below 100 GHz, several studies have reported deviations from the MBB extrapolation.
In the Large and Small Magellanic Clouds (LMC and SMC), an excess emission above the MBB prediction has been detected from the submillimeter to centimeter regime \citep{2010A&A...519A..67I,2011A&A...536A..17P}.
In contrast, such excess is not evident in M33 and M31, suggesting that the long-wavelength emission properties are determined by the dust composition, internal structure, and metallicity \citep{2014A&A...561A..95T}.

Several mechanisms have been proposed to explain the origin of the long-wavelength SED excess:
(1) enhanced absorption due to two-level systems in amorphous grains \citep{2007A&A...468..171M,2011A&A...534A.118P};
(2) spinning dust emission from rapidly rotating PAHs and nanoparticles, responsible for the so-called anomalous microwave emission observed at 10-60 GHz \citep{1998ApJ...494L..19D,2023ApJ...948...55H}; and
(3) magnetic dipole emission from magnetic grains \citep{2013ApJ...765..159D}.

In the cold dust disks within a few parsecs from the AGN, infrared spectra indicate an enhanced silicate fraction.
At several parsecs from the nucleus, the survival of small dust grains or PAHs may contribute to an excess continuum below 100 GHz, potentially through spinning-dust or magnetic-dipole processes. 
In particular, neutral PAH emission is observed in AGN environments
\citep[e.g.][]{2024A&A...691A.162G}.

\subsection{Circumnuclear Star Formation}
Circumnuclear starbursts (nuclear rings) within $\sim$10--100 pc of galactic centers produce radio emission powered by recent star formation. 
In the GHz band, non-thermal synchrotron emission from supernova remnants and diffuse cosmic-ray electrons typically dominates at lower frequencies \citep[see, e.g. ][]{1992ARA&A..30..575C}. 
Thermal free-free emission from H II regions contributes a flatter component at higher frequencies, and their superposition yields a broken power-law SED \citep{2011ApJ...737...67M}. 
The circumnuclear radio luminosity correlates with infrared emission, including reprocessed far-IR, and with optical SFR tracers such as H$\alpha$ and near-UV; because optical light suffers dust extinction, the IR-radio correlation is generally more robust where obscured star formation dominates \citep[e.g.][]{1985ApJ...298L...7H}. 

Improving spatial resolution from $\sim$100 pc to a few parsecs often changes the dominant radio source: extended star formation can dominate on $\sim$100 pc scales, whereas on a few pc scales compact components with flat-to-inverted spectra emerge, allowing spatial and spectral separation of AGN components \citep{2004ApJ...613..794G}. 
Thus, beam differences can confuse star formation and AGN contributions. Conversely, if the same target is imaged at matched frequencies but progressively finer beams, the disappearance of steep-spectrum extended emission and the emergence of flat/inverted compact emission provide direct evidence that different physical processes dominate different spatial scales. This is the observational logic that SKA can apply systematically to samples rather than only to a few bright nearby prototypes. 
In the nearby prototype NGC 1068, the core shows a compact $\sim$10-20 mJy component in the cm-submm bands, while a circumnuclear starburst ring at $\sim$100-300 pc can contribute comparable or greater low-frequency flux; spatially resolved SEDs and interferometry demonstrate the importance of separating components across both frequency and angular resolution axes \citep{2016ApJ...823L..12G,2019A&A...632A..61G,2018ApJ...853L..25I}.
High-resolution ALMA observations have similarly shown that, when the beam size decreases to the inner tens of parsecs, compact nuclear millimeter emission can be separated from circumnuclear star-forming components \citep{2016ApJ...826...59I}.
Recent ALMA studies of hard-X-ray-selected nearby AGN further show that nuclear millimeter luminosity correlates with intrinsic hard-X-ray luminosity, supporting the idea that the compact mm--radio component on small scales is closely linked to AGN activity rather than extended star formation \citep{2022ApJ...938...87K,2023ApJ...952L..28R}.

\section{Observational Feasibility with SKA}

\begin{figure*}[htbp]
\centering
\includegraphics[width=\textwidth]{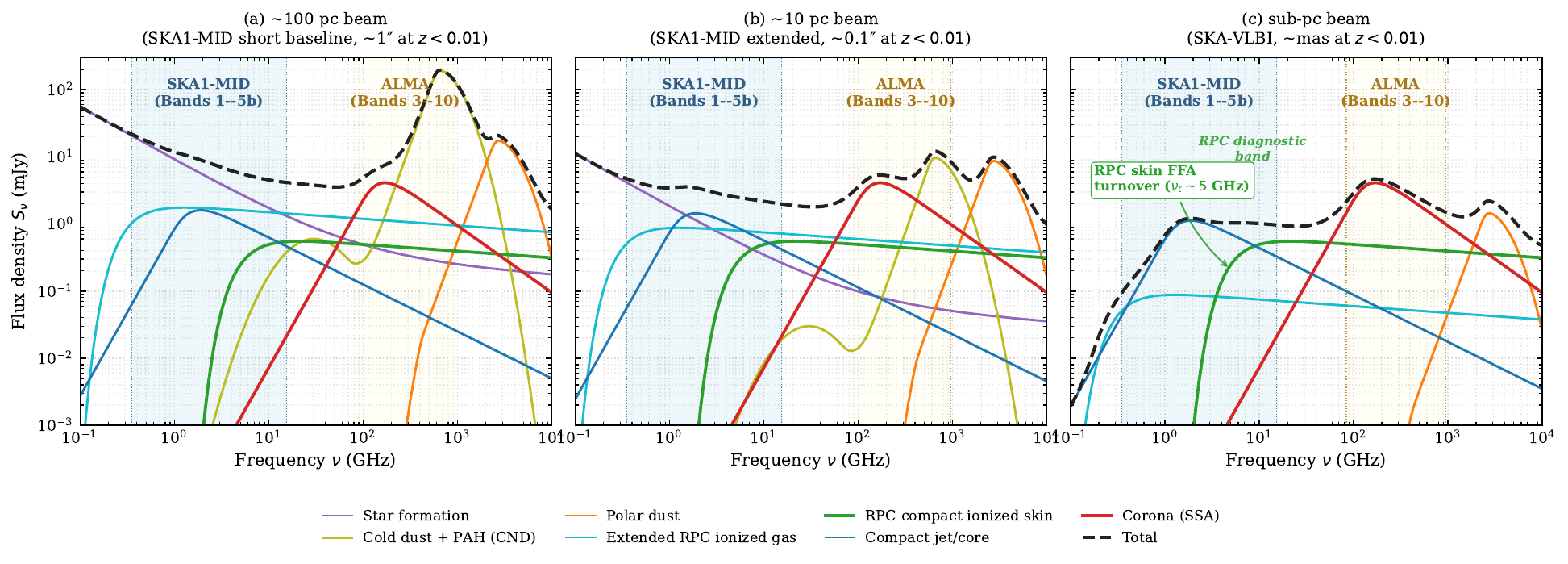}
\caption{
Beam-size-dependent SED of a representative nearby ($z<0.01$) Type 1-like radio-quiet AGN, modeled under the radiation-pressure-confinement (RPC) interpretation of the inner ionized gas (Section~2.3). 
Panels (a), (b), and (c) correspond to $\sim$100\,pc, $\sim$10\,pc, and sub-parsec beams, respectively. 
Extended components are progressively resolved out as the beam shrinks, while compact AGN-driven components persist. 
Vertical dotted lines mark the SKA1-MID frequency range and the ALMA Bands 3--10.
The compact RPC ionized skin shows an FFA turnover at $\nu_t \sim 5$\,GHz, inside the SKA Band 5b diagnostic range. 
Synchrotron self-absorption follows the T\"urler form, free-free absorption uses $\tau_{\rm ff}\propto\nu^{-2.1}$, and the spinning-dust bump is approximated by a log-Gaussian \citep{1998ApJ...494L..19D}. 
All component normalizations are illustrative.
}
\label{fig:sed_beam}
\end{figure*}

\begin{figure*}[htbp]
\centering
\includegraphics[width=\textwidth]{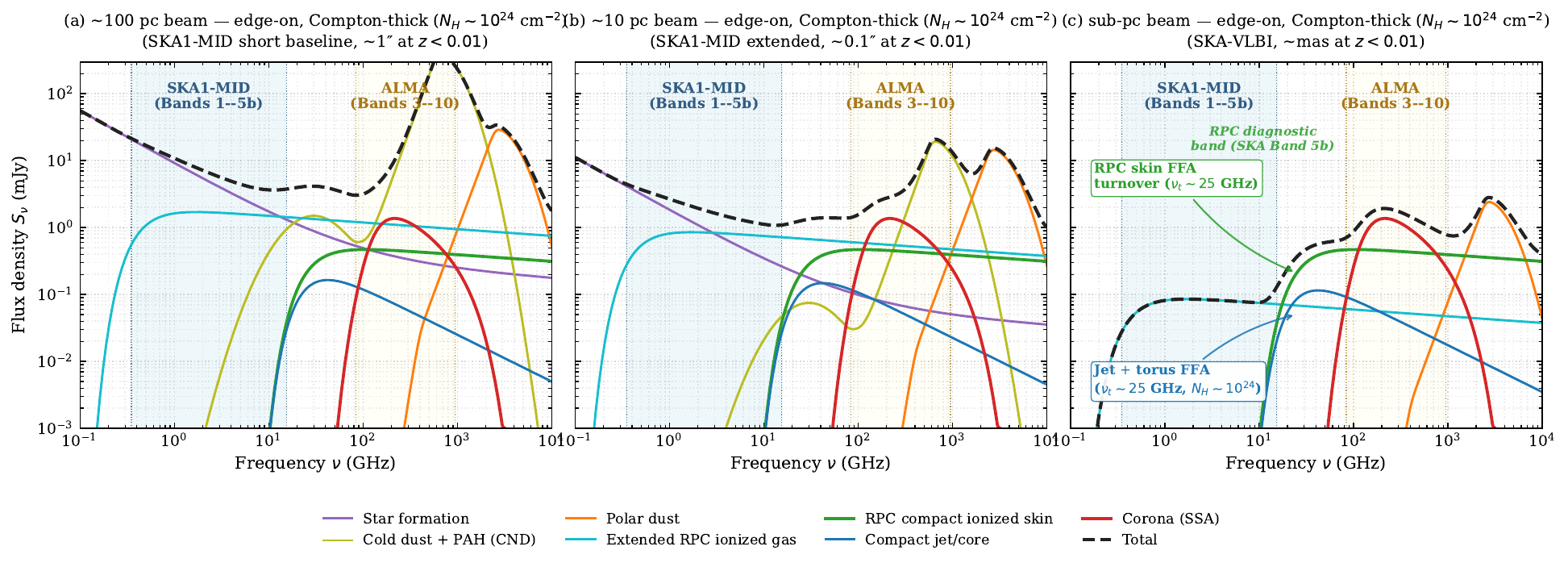}
\caption{
Same as Fig. 1, but for an edge-on, Compton-thick ($N_H \sim 10^{24}$ cm$^{-2}$) case. The main differences from the Type 1-like case are foreground free-free absorption of the inner jet/core, a shift of the RPC compact-skin turnover to $\nu_t \sim 25$ GHz, attenuation of the coronal SSA bump by combined torus FFA, dust extinction, and Compton scattering, and enhanced dust-related emission at high frequencies. These changes illustrate how viewing geometry modifies the SKA-band spectral diagnostics while leaving the high-frequency coronal SSA component accessible (though attenuated) to ALMA.
}
\label{fig:sed_beam_dusty}
\end{figure*}

We now translate the spectral and spatial diagnostics introduced in Section~2 into concrete predictions for SKA1-MID / AA$^{*}$ observations. The specific forecast is that spectral-index maps made at several controlled beam sizes will identify which radio-emission component dominates at each physical scale. In this sense, the spatial resolution is not simply a technical parameter but one of the axes of the science measurement.
Rather than enumerate every component again, we anchor the discussion in two synthetic spectral energy distributions that compactly summarize the expected behavior of all components introduced in Section~2: 
Fig.~\ref{fig:sed_beam} for a face-on / Type 1-like geometry, and Fig.~\ref{fig:sed_beam_dusty} for an edge-on / dust-dominated geometry such as a Compton-thick system. 
Throughout this section, we adopt nearby ($z \lesssim 0.01$, $d \lesssim 43$\,Mpc) RQ AGN as the representative target population, for which the angular-to-physical scale conversion is approximately 1\,mas\,$\sim$\,0.005--0.2\,pc and 1$''$\,$\sim$\,5--210\,pc.
This redshift cut is chosen so that structures smaller than $\sim$100\,pc, including the BLR region, polar dust, and dusty torus, can be spatially resolved with the SKA's milliarcsecond to sub-arcsecond beam.

\subsection{What the SED panels tell us about SKA observability}
\label{sec:sed_panels}

The two figures summarize the central observational argument: the components introduced in Section~2 are not only spectrally distinct, but also selected differently by the synthesized beam. 
They should therefore be read as forecasting diagrams for beam-dependent spectral-index science. 
The same intrinsic AGN can show different apparent $\alpha$ distributions because the beam mixes different fractions of circumnuclear star formation, ionized outflows, compact jets, coronal emission, and dust-related components. 
Thus, a change in spectral index with angular resolution is not merely an observational bias, but a diagnostic of the physical scale on which each emission process dominates.

Figures~\ref{fig:sed_beam} and~\ref{fig:sed_beam_dusty} illustrate two limiting geometries: a Type~1-like view and an edge-on, Compton-thick ($N_{\rm H} \sim 10^{24}$~cm$^{-2}$) view. 
Here we emphasize the observational implication. 
At $\sim$100~pc resolution, the measured radio SED is expected to be strongly affected by extended star formation, cold dust-related emission (including the PAH/spinning-dust bump near 30~GHz), and diffuse ionized gas. 
At $\sim$10~pc resolution, the relative contribution of compact AGN-related components increases as the cold dust and star formation contributions are progressively resolved out. 
At sub-parsec resolution, the SED is dominated by the compact jet/core, the RPC ionized skin, and the coronal component, with the cold dust contribution removed by the small beam area rather than by absorption. 
This progression provides the physical basis for using SKA spectral-index maps to identify where the dominant power source changes from host-galaxy star formation to nuclear activity.

The most important SKA diagnostic is the detection of 1--10~GHz spectral turnovers and their dependence on beam size and viewing geometry. 
In the Type~1-like case, the RPC compact-skin free--free absorption feature appears at $\nu_t \sim 5$~GHz, well inside SKA1-MID. 
In the edge-on, Compton-thick case, the same feature shifts to $\nu_t \sim 25$~GHz because of additional ionized material along the line of sight, placing it at the upper edge of SKA Band~5b. 
At the same time, free--free absorption from the obscuring torus suppresses the inner jet/core out to $\sim$25~GHz, producing a low-frequency cutoff that is itself a sensitive probe of the absorbing column density. 
Such turnover shifts provide a direct radio measure of orientation-dependent warm ionized columns, complementary to X-ray constraints on absorbing gas. 
Therefore, SKA Band~5 and Band~5b are particularly important for separating compact free--free emission/absorption from steep-spectrum synchrotron components, and for distinguishing Type~1-like from heavily obscured geometries through the location of the turnover.

A complementary diagnostic available at higher frequencies is the coronal SSA component near 100--300~GHz. 
In the Type~1-like geometry, the coronal SSA bump is directly accessible to ALMA. 
In the Compton-thick edge-on case, the same component is attenuated by a factor of $\sim$3 at 150~GHz through the combined effect of torus free--free absorption, foreground dust extinction, and Compton scattering ($\tau_{\rm T} \sim 0.7$ for $N_{\rm H} \sim 10^{24}$~cm$^{-2}$). 
The corona therefore remains observable through ALMA in both geometries, but the ratio between the observed and intrinsic coronal flux density provides an independent radio measure of the line-of-sight column density that can be cross-checked against X-ray $N_{\rm H}$ estimates.

These SED forecasts also define the observational strategy. 
First, matched-resolution imaging across SKA bands is required to construct reliable spectral-index maps without confusing intrinsic spectral curvature with beam dilution. 
Second, multi-scale imaging is required: low-resolution images recover extended star formation and diffuse outflows, while long-baseline SKA and SKA-VLBI observations isolate the compact nuclear components. 
Third, the SKA data should be interpreted together with ALMA observations at millimeter/submillimeter wavelengths and X-ray measurements of the ionized absorber. 
ALMA constrains the high-frequency dust and (attenuated) coronal components, while X-ray spectroscopy provides independent information on the ionization state and column density of the absorbing gas. 
The combined data set will therefore test whether the scale-dependent radio spectral-index structure is produced by the same multi-phase circumnuclear gas seen in infrared, optical, and X-ray diagnostics.

The principal SKA-band requirements following from these forecasts are broad frequency coverage from $\sim$1 to 15.4~GHz, sensitivity to compact components at the $\sim$0.1--1~mJy level, and angular resolution sufficient to separate $\sim$100~pc, $\sim$10~pc, and sub-parsec structures in nearby ($z<0.01$) AGN.
SKA1-MID provides the necessary sensitivity for both unobscured and Compton-thick cases, while SKA-VLBI will be essential for the closest sources where the sub-parsec structures can be resolved.

\subsection{Tiered observing strategy and connection to $\alpha$-mapping}

Resolved spectral-index mapping at centimeter wavelengths has a long pedigree in nearby AGN studies.
Multi-frequency VLA and VLBA observations of bright Seyferts have produced parsec-scale $\alpha$ maps that separate inverted/flat AGN cores from steep-spectrum jets and shock-driven outflow components \citep[e.g.,][]{2003ApJ...583..192M,2004ApJ...613..794G,2010MNRAS.401.2599O,2009AN....330..249S}, and recent e-MERLIN, EVN/VLBI, and JVLA studies have extended this to milliarcsecond and intermediate scales \citep{2021MNRAS.500.4749B,2023MNRAS.519.1732N,2021MNRAS.503.1780J,2023MNRAS.522.3506Z,2024ApJ...961..230S}.
However, current facilities leave a practical gap: the JVLA provides wide-band spectral imaging but at $\sim$0.2$''$ resolution at $\sim$10\,GHz (translating to $\sim$100\,pc only at $d \gtrsim 50$\,Mpc), whereas e-MERLIN/EVN/VLBI offer finer angular resolution but with narrower instantaneous frequency coverage and lower surface-brightness sensitivity.
SKA1-MID is therefore not simply a higher-resolution replacement: its key advantage is the combination of high sensitivity, broad 0.35--15.4\,GHz coverage, and multi-resolution imaging connecting extended circumnuclear emission to compact nuclear components.
Together with SKA-VLBI for the nearest sources, this enables systematic $\alpha$ and turnover-frequency mapping from $\sim$100\,pc to sub-parsec scales, identifying the spatial scale at which the dominant radio power source changes from host-galaxy star formation to nuclear activity.

\textit{Tier 1: SKA-only multi-scale SED decomposition and spectral-index mapping.}
SKA1-MID, supplemented by SKA-VLBI for the nearest targets, images each source across Bands~1--5b at a controlled set of matched angular resolutions, producing scale-dependent integrated SEDs and $\alpha$ maps at $\sim$100\,pc, $\sim$10\,pc, and sub-parsec resolutions.
Components that disappear as the beam shrinks are associated with extended star formation, diffuse outflows, or large-scale dust-related emission, whereas components that persist trace compact AGN structures, directly testing the logic of Figs.~\ref{fig:sed_beam} and~\ref{fig:sed_beam_dusty}.
At each fixed resolution, the diagnostic regimes follow directly from the SED morphology: $\alpha \approx -0.1$ indicates thermal free--free emission from photoionized gas; $\alpha \approx -0.7$ to $-1.0$ traces optically thin synchrotron from supernova remnants, outer jets, or shock-accelerated electrons; and $\alpha \gtrsim 0$ identifies optically thick or absorbed compact components such as self-absorbed cores, coronae, or FFA-affected jets.
The spatial variation of the peak frequency $\nu_{\rm peak}$ further distinguishes SSA from FFA, and Type~1-like from edge-on geometries, even when their global SEDs appear similar.
All maps must be produced after matching the $uv$ coverage and restoring beam between frequency pairs to avoid artificial gradients from extended-flux resolution effects.

\textit{Tier 2: Multifacility joint observations.}
Compact, spectrally interesting components identified in Tier~1 become targets for multi-facility follow-up (Section~4).
In brief: corona candidates with detected SSA tails in SKA Band~5b are followed up with ALMA Bands~3--7 to localize the SSA peak ($\nu_m$, $S_m$) and with XRISM/Athena for the inverse-Compton X-ray luminosity and ionization parameter $\xi$, jointly constraining $B$, $R$, $N_e$, and $U_{\rm ph}$.
RPC ionized-gas regions are followed up with VLT/MUSE and JWST/NIRSpec for kinematics and with X-ray spectroscopy for $\xi$ and $N_H$, providing a direct sub-pc test of the RPC framework.
Polar-dust candidates are followed up with VLTI/MATISSE and JWST/MIRI for direct dust thermal imaging at matched milliarcsecond resolution.

\textit{Tier 3: Time-domain monitoring.}
Coordinated multi-epoch SKA + ALMA + X-ray monitoring tests whether millimeter and X-ray flares are temporally correlated, establishing a causal connection between the SSA-emitting plasma and the inverse-Compton up-scattering electrons in the same coronal volume.
On longer timescales, repeated SKA imaging traces the dynamical evolution of jet ejections and outflow components, complementing the snapshot diagnostics of Tiers~1--2.

\section{Multi-frequency Synergy}

The SKA spectral-index and turnover maps described above become most powerful when anchored to independent tracers at other wavelengths.
Radio data identify where thermal, non-thermal, absorbed, and self-absorbed components are located, but they cannot by themselves determine the ionization state, dust temperature, gas velocity field, or X-ray coronal power.
Multi-frequency observations therefore provide the physical labels for the components isolated by SKA at each beam size.

At millimeter and submillimeter wavelengths, ALMA and the ngVLA connect the SKA band to the high-frequency emission from compact dust, molecular gas, and the coronal synchrotron component.
ALMA constrains the thermal dust and molecular torus on parsec scales, while the ngVLA will bridge the frequency gap between SKA and ALMA, especially around tens of GHz where compact nuclear components may turn over.
Together with SKA, these facilities separate steep-spectrum star formation, free--free emission, and compact SSA-dominated components.

X-ray observations provide the complementary view of the accretion-powered nucleus and ionized absorbers.
XRISM, Chandra, XMM-Newton, and future Athena observations measure the coronal luminosity, ionization parameter, column density, and outflow velocity, which are needed to distinguish coronal synchrotron emission from free--free emission or absorption by ionized gas.
This is particularly important for testing whether the SKA-detected compact ionized component follows the RPC-like conditions inferred from X-ray spectroscopy.

Infrared facilities such as JWST, VLTI/GRAVITY+, VLTI/MATISSE, and ELT/HARMONI trace the dusty and ionized structures that are unresolved or only indirectly detected in the radio.
JWST/MIRI maps high-ionization mid-infrared lines and warm polar dust on tens-to-hundreds of parsec scales, while VLTI and ELT observations probe the dust sublimation region and BLR--NLR transition on parsec to sub-parsec scales.
These data test whether the compact radio free--free component is associated with the same dusty ionized outflow seen in infrared interferometry and recombination-line kinematics.

Optical IFU observations, especially with VLT/MUSE and future facilities, provide the large-scale ionized-gas kinematics through [O\,{\sc iii}], H$\alpha$, [N\,{\sc ii}], and [S\,{\sc ii}] emission.
Comparing these maps with SKA free--free, synchrotron, and polarization/rotation-measure maps distinguishes photoionized gas from shock-accelerated outflow components and tests whether magnetic fields contribute to launching or collimating the wind.

Thus, the unifying observable is the scale dependence of the radio spectral index.
A steep $\alpha$ measured with a $\sim$100\,pc beam can trace star formation or extended outflow shocks, whereas a flat or inverted $\alpha$ emerging only at $\sim$10\,pc or sub-pc resolution indicates compact jet emission, FFA by dense ionized gas, or coronal SSA.
By tying these scale-dependent radio components to ALMA/ngVLA, X-ray, JWST/VLTI, and optical-IFU diagnostics, SKA will enable a simultaneous view of circumnuclear star formation, dusty ionized outflows, torus-scale gas, and the accretion-powered compact nucleus in radio-quiet AGN.

\begin{table}[h]
\centering
\caption{Multi-frequency diagnostics for radio-quiet AGN nuclear components. Columns indicate the SKA-only diagnostic, the complementary observation needed to break degeneracies, and the corresponding facility (current or planned).}
\label{tab:synergy}
\small
\setlength{\tabcolsep}{4pt}
\renewcommand{\arraystretch}{1.15}
\begin{tabular}{@{}p{0.16\linewidth}|p{0.24\linewidth}|p{0.30\linewidth}|p{0.22\linewidth}@{}}
\hline
\textbf{Component} & \textbf{SKA diagnostic} & \textbf{Complementary observation} & \textbf{Facility} \\
\hline
Compact jet / core
& Spectral index $\alpha$, brightness temperature $T_b$, core shift, polarization
& mm-band SED, SSA turnover, $\gamma$-ray/neutrino identification
& ALMA, ngVLA, CTA, IceCube \\
\hline
Disk corona (SSA)
& Optically thick rising spectrum, low-$\nu$ tail
& mm SSA turnover ($\nu_m$, $S_m$), X-ray IC luminosity
& ALMA Band 3--7, ngVLA, XRISM, Athena \\
\hline
Polar dust / inner torus
& cm/mm free-free contribution from photoionized skin
& Direct dust imaging, 10~$\mu$m silicate feature, K/N-band interferometry
& JWST/MIRI, VLTI/GRAVITY+/MATISSE \\
\hline
Ionized gas outflow (NLR)
& GHz free-free flux, spectral index $\alpha\sim-0.1$
& Ionization parameter $\xi$, $N_H$, outflow kinematics, [O~III]/H$\alpha$ kinematics
& XRISM, Athena/X-IFU, VLT/MUSE, JWST/NIRSpec \\
\hline
Cold dust in CND
& cm/mm long-wavelength SED excess (anomalous microwave)
& far-IR/submm SED peak, dust composition (silicate/PAH)
& ALMA Band 7--10, JWST/MIRI, Herschel archive \\
\hline
Circumnuclear star formation
& Non-thermal synchrotron + thermal free-free, $\sim$100~pc structure
& IR luminosity, H$\alpha$/Br$\gamma$ SFR, supernova rates
& JWST, ALMA, MUSE, Chandra \\
\hline
Magnetic field structure
& Faraday RM synthesis, polarization fraction
& Optical/IR polarimetry, MHD simulations
& VLT/SPHERE, ELT \\
\hline
\end{tabular}
\end{table}

\section*{Acknowledgments}
We thank the reviewer for this helpful suggestion. 
Y.K. was supported by JSPS KAKENHI Grant Number 24K17080.
This work is partially supported by MEXT as “Project for Establishment of a Center for Advanced HPC-AI Development Support”.
Y.K. used AI language models for language editing and figure generation.
The authors take full responsibility for the scientific content of this chapter.


\bibliographystyle{abbrvnat-maxbibnames4}
\bibliography{chapter} 

\end{document}